\documentstyle[12pt]{article}
\begin{document}
\baselineskip 0.3in

\begin{titlepage}
\thispagestyle{empty}
\hfill
\begin{center}
{\Large \bf
An exactly solvable lattice model for inhomogeneous interface growth
}\\[15mm]

{\large {\sc
Gunter M. Sch\"utz}
} \\[5mm]

\begin{center}
{\small\sl
Department of Physics, University of Oxford,\\
Theoretical Physics, 1 Keble Road, Oxford OX1 3NP, UK\\ 
and\\
Laboratoire de Physique du Solide\footnote{Unit\'e de recherche 
associ\'ee au CRNS no. 155}, 
Universit\'e Henri Poincar\'e, \\
B.P. 239, F-54506 Vand{\oe}uvre-l\`es-Nancy Cedex, France\\
e-mail: schutz@vax.ox.ac.uk
}
\end{center}
\vspace{10mm}
\end{center}
{\small
We study the dynamics of an exactly solvable lattice model for inhomogeneous 
interface growth. The interface grows deterministically with constant velocity 
except along a defect line where the growth process is random. We obtain exact 
expressions for the average height and height fluctuations as functions of space 
and time for an initially flat interface. For a given defect strength there is
a critical angle between the defect line and the growth direction above which
a cusp in the interface develops. In the mapping to polymers in random media
this is an example for the transverse Meissner effect. Fluctuations around 
the mean shape of the interface are Gaussian.
}\\
\vspace{2mm}
\end{titlepage}
\newpage

The problem of nonlinear, KPZ-type interface growth in 1+1 dimensions and
the closely related problem of polymers in random media has been the subject of
many studies over the past decade \cite{KS}. The underlying assumption in this 
approach is that
growth occurs stochastically normal to the interface. The dynamics of the
interface are then given by the KPZ equation $\partial_t h = \nu \nabla^2 h + 
\lambda (\nabla h)^2 + \eta$ \cite{KPZ}. The deterministic version of the 
equation (without the noise $\eta$) can be solved explicitly. Among other
things one finds that in the infinite 
time limit the interface becomes flat. However, in the
presence of noise, even an initially flat interface roughens and the derivation
of quantities related to the roughening process and other fluctuation phenomena 
become an interesting and challenging problem, usually tackled using either
the renormalization group or the study of lattice models. In this context the 
asymmetric 
exclusion process \cite{SZ}, a driven lattice gas model, has played a special
role. Both numerical and exact analytical
results have been obtained from this microscopically motivated exactly solvable 
model, which maps to a lattice growth model in the 
universality class of the noisy KPZ equation.

Recently Kallabis and L\"assig studied the KPZ equation with spatially
localized noise and found an interesting phase diagram even in one dimension
\cite{KL}. While this work focussed on the properties of the directed polymer
mapping, subsequent work elucidated some properties of the growth model itself
\cite{NK}. Among other things it was found that in one dimension the interface
in the steady state has a cusp with constant slope for arbitrarily small noise. 
This observation is, in fact, not new: Already some while ago, an exclusion
process on a lattice was introduced which mimics the situation of homogeneous 
and deterministic growth except along one defect line where growth is 
stochastic \cite{G}. In the steady state of this non-equilibrium system one 
finds indeed such a cusp, again for arbitrarily small noise if the 
defect line is parallel to the average growth direction as assumed in the 
continuum model studied in Refs. \cite{KL,NK}. This makes a study of 
the dynamical properties of the lattice model desirable. It is the aim of this 
paper to study the time evolution of an initially flat interface using this
model.

The model introduced in Ref. \cite{G} is an exactly solvable version of the 
deterministic asymmetric exclusion process on a ring with a defect line. We use 
the well-known RSOS mapping from an interface to a lattice gas \cite{MRSB,PRL}: 
In the interface model the (integer) height variables $h(y,t)$ may differ on
neighbouring (integer) sites $y,y+1$ only by $\pm 1$. As a result of this
restriction, growth can occur only in local minima, no overhangs can develop.
If a growth event occurs, the height at site $y$ increases by two units
(see Fig. 1). In lattice gas language
the height differences between neighbouring sites are mapped to 
a particle occupation number $n_x(t)=0,1$ with the presence of a particle
on $x$ corresponding to slope -1 between sites $y-1$ and $y$ in the interface
model and a vacancy at site $x$ corresponding to slope +1 (see Fig. 1).
Growth at site $y$ in the growth model then corresponds to a particle hopping
from site $x$ to a vacant site $x+1$. Note that each lattice site can be
occupied by at most one particle. In a finite system of $L=2N$ sites for the
particle model one needs $N$ particles in order to ensure periodic
boundary conditions in the height model. In an infinite system, an average
density $\rho \neq 1/2$ over some finite range $L'$ corresponds to a
non-zero average slope $1-2\rho$ in that range.

So far this mapping keeps track only of the height differences between 
neighbouring sites. In order to make the two models equivalent, one has to 
introduce an additional integer random variable $h_0 \in {\bf Z}$ in the 
particle system which gives the absolute height of the
interface at some arbitrary, but specified point, e.g. $y=0$. In the
lattice gas mapping the value of this random variable is increased by
two units each time a particle hops from site 0 to site 1. From this
one may reproduce the height at any point since $h(y) = h_0 + \sum_{x=1}^y
(1-2 n_x)$.

The deterministic time evolution is realized by a parallel updating scheme
in which in a first step all odd pairs of sites $(2x-1,2x)$ are updated
according to the following rules: If there is a particle on site $2x-1$
and a vacancy on site $2x$, then the particles hop with probability 1 to 
site $2x$. If the pair of sites is in one of the three remaining configurations,
nothing changes. These rules are applied in parallel to all such pairs. In a 
next step one shifts the pairing by one lattice unit and applies the same rules 
to the even pairs $(2x,2x+1)$. This completes one full time step. Note that so 
far this model is fully deterministic, and any initial configuration will 
eventually evolve into a flat interface characterized by a particle 
configuration $(\dots, 1,0,1,0,1,0,1,0, \dots)$. The dynamical properties of a 
similar system with slightly different deterministic updating rules were 
studied in \cite{KSxx}.

In order to introduce a defect line at $y=0$ we change the rules 
such that a particle on site $x=0$ hops only with probability
$p \leq 1$, if site 1 is vacant. This randomness results in non-trivial
behaviour of the system. The stationary properties of this model on a finite 
ring with $N$ particles were studied in \cite{G} ($N$ arbitrary). Here we
shall consider the dynamics of the infinite system with flat initial
configuration  $(\dots, 1,0,1,0,1,0,1,0, \dots)$ where the particles are
on the odd sites while the vacancies are on the even sites. Because of
reflection symmetry one has $h(y) = h(-y)$.

It is useful to note that the time evolution may be written in terms of
a transfer matrix $T = p T_0 + (1-p) T_1$ acting on states corresponding to
some given configuration of the system \cite{G}. Any state is fully 
characterized by the particle occupation numbers $n_x$ and the integer height 
variable $h_0$. $T_0$ is the transfer matrix for the system without defect, 
while $T_1$ is the transfer matrix of the system with full blockage where a 
particle at site 0 can never move to site 1, even if it is vacant. Acting with 
$T_0$ on the state $(\dots, 1,0,1,0,1,0,1,0, \dots)$ reproduces this state by 
shifting the whole configuration by two lattice units to the right. This is in 
accordance with the remark that this configuration is stationary if $p=1$. The 
height variable $h_0$ increases by two units since a particle went from site 0 
two site 1. This means that the interface has grown everywhere by two units. On 
the other hand, the action of $T_1$ on this state results in a configuration
$(\dots , 0,0,1,0,1,0,1,0 \dots)$ where the first occupation number in this 
set (which is 0) corresponds to the occupation on site 1. The occupation 
numbers on site 0 and on the negative sites follow from symmetry. In this case 
$h_0$ has not increased, only $h(y)$ for $y \leq 1$ has grown by two units. 
Taking the $t^{th}$ power of $T$ yields all possible configurations the 
interface may take after $t$ time steps. Each realization acquires a factor
$p^k(1-p)^{t-k}$ which gives the probability that this particular growth history
has taken place.

Using the decomposition of $T$ into $T_0$ and $T_1$ it becomes straightforward 
to calculate to respective probabilities for all these configurations: (1) The 
odd sublattice remains empty for all times. (2) The consequence of the defect 
for the even positive sublattice is the injection of a particle on site 0 with
probability $p$ in each time step. (3) If a particle has been injected it will 
move in each subsequent time step by two lattice units without any interaction
with other particles. As an example for what this corresponds to in height
language, consider $h_0$ which we take
to be zero at time $t=0$. With the dynamical rules as explained above one
finds after one step $h_0=1$ with probability $p$ and $h_0=0$ with probability
$1-p$. After two steps one finds $h_0=2$ with probability $p^2$, $h_0=1$ with 
probability $2p(1-p)$ and $h_0=0$ with probability $(1-p)^2$, and so on.
By construction, the height on the even sublattice will always be even, while
on the odd sublattice it will always be odd. Let us denote $h(2y,t) = 2 
\tilde{h}(2y,t)$. Then one gets for the probability $P_{y,t}(h)$ of finding the 
height $h$ at site $y$ at time $t$ (for $y \geq 0$)
\begin{eqnarray}
\label{1}
P_{2y,t}(h) & = & 
\mbox{$\left( \begin{array}{c} t-y \\ \tilde{h}-y \end{array} \right)$}
(1-p)^{t-\tilde{h}} p^{\tilde{h}-y} 
\hspace{4mm} (0 \leq y \leq t) \nonumber \\
 & = & \delta_{\tilde{h},t} \hspace{4mm} ( y \geq t)
\end{eqnarray}
On the odd (positive) lattice sites one has always, i.e. for all times and all 
realizations of the randomness, $h(2y-1,t) = h(2y,t)-1$. For the negative half 
space $y<0$ one gets $P[h(-y,t)] = P[h(y,t)]$ by symmetry. This result gives a 
complete description of the time evolution of the local heights of an interface 
which is initially (macroscopically) flat.\footnote{Eq. (1) is easy to verify 
e.g. by applying $T$ two or three times to the initial state. This makes the 
structure of the interface after $t$ steps already quite clear.} Note that 
there are no height fluctuations at all for $y \geq t$. This must be so since
the perturbation of the interface caused by the defect spreads with finite
velocity $v=1$ which is the bulk growth velocity.

In what follows we study only the non-negative even sublattice up to site
$y=t$, since all quantities relating the odd sublattice, the negative half 
space and the region $y > t$ are then trivially given. From the distribution 
(\ref{1}) one obtains the generating
function for height fluctuations
\begin{eqnarray}
\label{2}
\langle e^{\alpha h(2y,t)} \rangle & = & e^{2\alpha y} 
(1-p + p e^{2\alpha})^{t-y}
\hspace{4mm} (0 \leq y \leq t) \nonumber \\
 & = &  e^{2\alpha t} \hspace{4mm} ( y \geq t)
\end{eqnarray}
and in particular the average height 
\begin{eqnarray}
\label{3}
\langle h(2y,t) \rangle & = & 2t - 2(1-p)(t-y) \hspace{4mm} (0 \leq y \leq t) 
\nonumber \\
 & = &  2t \hspace{4mm} ( y \geq t).
\end{eqnarray}
The height fluctuations can be obtained by taking the second logarithmic
derivative of the generating function. One finds
\begin{eqnarray}
\label{4}
\langle h^2(2y,t) \rangle - \langle h(2y,t) \rangle^2 & = & 
4 p(1-p)(t-y) \hspace{4mm} (0 \leq y \leq t) \nonumber \\
 & = &  0 \hspace{4mm} ( y \geq t)
\end{eqnarray}
From Eq. (\ref{3}) one finds that the defect causes
the development of a cusp with constant slope $2(1-p)$. At site $y=0$ the
interface grows with constant speed $v_0 = 2p$, while in the bulk, at distances
larger than t, it grows with velocity $v_b = 2$ (Fig. 2). From (\ref{1}) one 
finds that the fluctuations of the interface round its mean value (\ref{3}) are
Gaussian (for $t$ large) with a space-dependent variance given by (\ref{4}).
These are the main results of this paper.

We conclude by briefly discussing the situation in which the defect
line is not parallel to the growth direction of the interface. This can be
realized within this model by taking a (macroscopically) non-zero initial
slope $s=1-2\rho$ of the interface. Since growth occurs normal to the interface 
the net result is indeed a defect line tilted relative to the main growth
direction. In particle language one has to
choose an initial state with a density $\rho \neq 1/2$., e.g. a configuration
$(\dots, 1,0,0,0,1,0,0,0, \dots)$. From the analysis of the steady state
of this system one finds that a phase transition takes place at $p_c = 2\rho$
for $\rho \leq 1/2$ (positive initial slope) and $p_c = 2(1-\rho)$
for $\rho \geq 1/2$ (negative initial slope) respectively \cite{G} . 
For the dynamics of an initially flat interface the steady state properties
of the system suggest that for 
$p > p_c$ the interface remains flat for all times except in a finite region
close to the defect. For $p<p_c$ a cusp will evolve. Unlike in the situation
described above, the boundary of this cusp will not be sharp, but fluctuate.
In a finite system these fluctuations scale in the system size like $L^{1/2}$
and one would therefore expect a power growth $t^{1/2}$ in time of these domain 
boundary 
fluctuations in the infinite system. For fixed $p$ this is a phase transition
that takes place at critical slopes $s_c = \pm (1-p)$. For $-(1-p) \leq s \leq
(1-p)$ the system develops the cusp.

This situation is also of interest in the well-known mapping of this problem to 
a directed polymer in a medium with a random defect line, as studied in Ref. 
\cite{KL} for the special case $\rho=1/2$ where the defect line and the average 
direction of the polymer are parallel. The development of the cusp corresponds 
to having a bound state where the polymer is bound to the
(attractive) defect line. It is no surprise that this happens at arbitrarily 
small defect strength. This simply reflects the fact that in one dimension a 
delta function potential (the defect line) with arbitrarily small amplitude
has a bound state. Translated into polymer 
language the phase transition in the interface model is a transition from an 
bound state of the polymer for a tilt of the defect line against the 
(average) direction of the polymer below the critical angle $s_c$ to an
unbound state for a tilt above $s_c$. This is a simple example for the 
occurrence of the transverse Meissner effect \cite{NV}.

A macroscopic cusp with constant slope appears to be an universal feature of 
growth models with a defect line where the growth process is impeded. In this
sense some features of the model studied in this paper may be of more than
purely theoretical interest. Normal growth, which is the underlying assumption
that leads to the KPZ equation, has been reported to occur e.g. in experiments
on wetting of tapes plunged in a liquid \cite{Blake}. It would therefore be 
interesting to adapt the model to this situation by an appropriate choice
of boundary conditions. On the other hand, our simple model
has some clear limitations. The strict discontinuity of the slope of the
interface at $y=t$ (for the half-filled system) is obviously an artefact of
the dynamics chosen here and cannot be expected to be found in any real system.
One would rather expect a smeared out discontinuity as in the system with a 
tilted defect or as in the fully stochastic version of the model with a defect 
\cite{TW}. Another, intrinsic, limitation of the model is that growth at the
defect is always slower than in the bulk. It would be very interesting to
study a model where the defect growth could be made faster than the
bulk growth velocity. 

The author would like to thank M. L\"assig and H. Kallabis for 
useful discussions. This work was supported by an EC Fellowship under the
Human Capital and Mobility program.

\newpage

\bibliographystyle{unsrt}

\newpage
\noindent
{\bf List of Figure Captions}

\vskip 1.in
\noindent
Fig. 1: 
The mapping between the restricted interface and the particle exclusion
process: we show a possible interface configuration and the corresponding 
particle occupancies on a lattice with sites labeled by $y$.
The indicated flips in the 
interface correspond to particles hopping on the lattice, marked by
horizontal arrows.
\vskip 1.in
\noindent
Fig. 2:
Macroscopic average interface shape at time $t=0$ (flat interface with height 0)
and at some later time $t$. The position of the defect at $y=0$ is indicated 
by the vertical dashed line. The defect line is parallel to the bulk growth
direction.

\newpage

\begin{figure}
\setlength{\unitlength}{1.0mm}
\begin{center}
\begin{picture}(100,80)
\put (-5,9){$h=0$}
\put (-5,59){$h=5$}
\put (45,0){$y$}
\multiput (10,9)(10,0){9}{\line(0,1){2.}}
\multiput (10,60)(1,0){85}{\circle*{0.1}}
\put (10,10){\line(1,0){8.}}
\put (15,10){\circle*{2}}
\put (16,10){\line(1,0){8.}}
\put (25,10){\circle{2}}
\put (26,10){\line(1,0){8.}}
\put (35,10){\circle{2}}
\put (36,10){\line(1,0){8.}}
\put (45,10){\circle*{2}}
\put (46,10){\line(1,0){8.}}
\put (55,10){\circle{2}}
\put (56,10){\line(1,0){8.}}
\put (65,10){\circle*{2}}
\put (66,10){\line(1,0){8.}}
\put (75,10){\circle*{2}}
\put (76,10){\line(1,0){8.}}
\put (85,10){\circle{2}}
\put (86,10){\line(1,0){8.}}
\multiput (10,12.5)(0.0,5.0){8}{\line(0,1){2.5}}
\multiput (20,12.5)(0.0,5.0){6}{\line(0,1){2.5}}
\multiput (30,12.5)(0.0,5.0){8}{\line(0,1){2.5}}
\multiput (40,12.5)(0.0,5.0){10}{\line(0,1){2.5}}
\multiput (50,12.5)(0.0,5.0){8}{\line(0,1){2.5}}
\multiput (60,12.5)(0.0,5.0){6}{\line(0,1){2.5}}
\multiput (70,12.5)(0.0,5.0){8}{\line(0,1){2.5}}
\multiput (80,12.5)(0.0,5.0){6}{\line(0,1){2.5}}
\multiput (90,12.5)(0.0,5.0){8}{\line(0,1){2.5}}
\thicklines
\put (18,6){\vector(1,0){4.0}}
\put (62,6){\vector(-1,0){4.0}}
\put (20,50){\vector(0,1){4.0}}
\put (60,50){\vector(0,-1){4.0}}
\put (10,50){\line(1,-1){10.}}
\put (20,40){\line(1,1){20.}}
\put (40,60){\line(1,-1){10.}}
\put (50,50){\line(1,1){10.}}
\put (60,60){\line(1,-1){20.}}
\put (80,40){\line(1,1){10.}}
\multiput (10,50)(1,1){10}{\circle*{0.1}}
\multiput (20,60)(1,-1){10}{\circle*{0.1}}
\multiput (50,50)(1,-1){10}{\circle*{0.1}}
\multiput (60,40)(1,1){10}{\circle*{0.1}}
\end{picture}
\end{center}
\caption{ }
\end{figure}

\begin{figure}
\setlength{\unitlength}{1.0mm}
\begin{center}
\begin{picture}(100,80)

\put (90,63){$h(y,t)$}
\put (90,12){$h(y,0)$}
\put (-5,59){$h=2t$}
\put (-5,49){$h=2pt$}
\put (-5,9){$h=0$}
\put (50,0){$y$}

\multiput (55,8.)(0.0,5.0){12}{\line(0,1){2.5}}

\thicklines
\put (10,9){\line(1,0){90.}}
\put (10,60){\line(1,0){25.}}
\put (35,60){\line(2,-1){20.}}
\put (55,50){\line(2,1){20.}}
\put (75,60){\line(1,0){25.}}

\end{picture}
\end{center}
\caption{ }
\end{figure}

\end{document}